\documentstyle[12pt]{article}
\begin{document}
\title{Nucleon Polarizabilities}
\author{Barry R. Holstein\\
Department of Physics-LGRT\\
University of Massachusetts\\
Amherst, MA 01003}
\maketitle
\begin{abstract}
The subject of nucleon polarizabilities in real, 
virtual, and doubly virtual Compton scattering is discussed 
with respect to what is known and how such quantitities can 
be extracted from data.
\end{abstract}
\newpage

\section{Introduction}

The subject of nucleon polarizabilities---both real and virtual---has 
become a hot one lately with activity on both the experimental and
theoretical fronts.  In this article, we present a brief review
of some of the interesting issues, including the latest experimental results
as well as new developments on the theoretical front, wherein dispersion
relations allow extraction of ``higher order'' nucleon
polarizabilities as well as connections to be made between the GDH sum
rule for real photons and the Bjorken sum rule for high $Q^2$.

\section{RCS: Ordinary Polarizabilities}

Historically, the basic idea of polarizabilities and their meaning
comes from the idea of the response of a system to the application of
an external quasistatic electromagnetic field\cite{ho}.  For example, in the
presence of an external electric field $\vec{E}_0$ a system of charges
will in general deform, with positive charges moving one way and negative
charges the other, resulting in an induced dipole moment $\vec{p}$
whose size is proportional to the strength of the applied field
\begin{equation}
\vec{p}=4\pi\alpha_E\vec{E}_0
\end{equation} 
The electric polarizability $\alpha_E$ is then the constant of
proportionality between the applied field and the induced dipole
moment.  Similarly one can define the magnetic polarizability $\beta_M$
in terms of the induced magnetic dipole moment $\vec{\mu}$ in the 
presence of an external mmagnetizing field $\vec{H}_0$
\begin{equation}
\vec{\mu}=4\pi\beta_M\vec{H}_0
\end{equation} 
For a macroscopic system it is intuitively clear how to measure such 
properties, but for an elementary particle, one must generate the electric
and magnetic fields via photons and use Compton scattering in order to
provide quantification.  In order to see how this is done first consider
RCS at very low energy---say $\omega << 20$ MeV---wherein the photon 
wavelength is much
longer than the size of the nucleon.  In this case, one is unable
to resolve the structure of the target and is sensitive only to its 
overall charge--$e$--and mass--$m.$  The interaction is then described 
by the simple lowest order Hamiltonian
\begin{equation}
H={(\vec{p}-e\vec{A})^2\over 2m}+e\phi
\end{equation}
and the resultant Compton scattering amplitude has the canonical Thomson form
\begin{equation}
{\rm Amp}=-{e^2\over m}\hat{\epsilon}'\cdot\hat{\epsilon}
\end{equation}
At higher energy---shorter wavelength---the internal structure becomes visible and
one can describe the interaction in terms of an effective Hamiltonian
having certain elementary properties---
\begin{itemize}
\item [i)] quadratic in $\vec{A}$;
\item [ii)] gauge invariant;
\item [iii)] rotational scalar;
 \item [iv)] P,T even, etc.
\end{itemize}
To the next leading order then the resultant form of the interaction 
is unique and must have the from
\begin{equation}
H_{eff}=-{1\over 2}4\pi\alpha_E\vec{E}^2-{1\over 2}4\pi\beta_M\vec{H}^2
\end{equation}
where $\alpha_E,\,\beta_M$ are just the electric, magnetic polarizabilities
defined above, as can be seen from the definitions
\begin{equation}
\vec{p}=-{\delta H_{eff}\over \delta\vec{E}}=4\pi\alpha_E\vec{E};\quad
\vec{\mu}=-{\delta H_{eff}\over \delta\vec{H}}=4\pi\beta_M\vec{H}
\end{equation}
Using the effective Hamiltonians given above the
Compton scattering amplitude becomes
\begin{equation}
{\rm Amp}^{(2)}=\hat{\epsilon}\cdot\hat{\epsilon}'\left({-e^2\over
M}+\omega\omega' \; 4\pi\alpha_E^p \right)
+\hat{\epsilon}\times\vec{k}\cdot\hat{\epsilon}'
\times\vec{k}'\;4\pi\beta_M^p+\; {\cal O}(\omega^4) \; .
\end{equation}
and the resultant differential scattering cross section is
\begin{eqnarray}
{d\sigma\over d\Omega}&=&\left({\alpha\over M}\right)^2
\left({\omega'\over \omega}\right)^2\left[{1\over 2}
(1+\cos^2\theta)\right.\nonumber\\
&-&\left.{ M\omega\omega'\over \alpha}\left({1\over 2}
(\alpha_E^p+\beta_M^p)(1+\cos\theta)^2+{1\over
2}(\alpha_E^p-\beta_M^p)(1-\cos\theta)^2\right)+\ldots\right],
\nonumber\\
\quad
\end{eqnarray}
here $\alpha=e^2/4\pi$ is the fine structure constant.  It is clear 
then that
$\alpha_E,\,\beta_M$ can be extracted via careful measurement of the
differential cross section and previous experiements at SAL and MAMI have
yielded the values\cite{nat}\footnote{In order to put these
numbers in perspective
note that for a hydrogen atom one finds $\alpha_E(H)\sim$ Volume($H$)
while for
the proton Eq. \ref{eq:yy} gives $\alpha_E(p)\sim 10^{-3}$Volume($p$), so that
the proton is a much more strongly bound system.}
\begin{equation}
\alpha_E^p=(12.1\pm 0.8\pm 0.5)\times
10^{-4}\,\,{\rm fm}^3;\quad\beta_M^p=(2.1\mp 0.8\mp 0.5)\times
10^{-4}\,\,{\rm fm}^3.\label{eq:yy}
\end{equation}
Recently preliminary new values have been announced obtained from precise
$p(\gamma,\gamma)p$ measurements using the TAPS and LARA spectrometers
\cite{wis}
\begin{equation}
\alpha_E^p=(12.24\pm 0.24\pm 0.54)\times
10^{-4}\,\,{\rm fm}^3;\quad\beta_M^p=(1.57\mp 0.24\mp 0.54)\times
10^{-4}\,\,{\rm fm}^3.
\end{equation}
Hemmert will show how these numbers compare with values obtained via
chiral perturbation theory.

The above results are well known and our task today is to extend
this discussion
to indlude spin degrees of freedom.  In this case the general Compton
amplitude can written in the general form
\begin{eqnarray}
T &=& A_1(\omega,z)\vec{\epsilon}^{\, \prime}\cdot\vec{\epsilon}
+A_2(\omega,z)\vec{\epsilon}^{\,  \prime}\cdot\hat{k} \; \vec{\epsilon}
\cdot\hat{k}^\prime \nonumber\\
&+&iA_3(\omega,z)\vec{\sigma}\cdot(\vec{\epsilon}^{\, \prime}\times
\vec{\epsilon})
+iA_4(\omega,z)\vec{\sigma}\cdot(\hat{k}^\prime \times\hat{k})
\vec{\epsilon}^{\,  \prime} \cdot\vec{\epsilon} \nonumber\\
&+& iA_5(\omega,z)\vec{\sigma}\cdot[(\vec{\epsilon}^{\,  \prime} \times
\hat{k}) \vec{\epsilon}\cdot\hat{k}^\prime -(\vec{\epsilon}\times
\hat{k}^\prime ) \vec{\epsilon}^{\,  \prime} \cdot\hat{k}]\nonumber\\
&+& iA_6(\omega,z)\vec{\sigma}\cdot[(\vec{\epsilon}^{\, \prime}\times
\hat{k}^\prime ) \hat{\epsilon}\cdot\hat{k}^\prime -(\vec{\epsilon}\times
\hat{k})\vec{\epsilon}^{\, \prime} \cdot\hat{k}],
\end{eqnarray}
and each amplitude can be expanded in terms of a lowest order Born
contribution plus a higher order and structure dependent polarizability
term.  In the case of the spin-dependent amplitudes $A_{3,4,5,6}$ such
structure effects arise at ${\cal O}(\omega^3)$ and can be characterized in
terms of an effective Hamiltonian involving four "spin-polarizabilities"
\begin{equation}
H_{eff}^{(3)}=-{1\over 2}4\pi(\gamma_{E1}^p\vec{\sigma}\cdot\vec{E}\times
\dot{\vec{E}}+\gamma_{M1}^p\vec{\sigma}\cdot\vec{H}\times
\dot{\vec{H}}-2\gamma_{E2}^pE_{ij}\sigma_iH_j+2\gamma_{M2}^pH_{ij}\sigma_iE_j)
\label{eq:mno}
\end{equation}
where
\begin{equation}
E_{ij}={1\over 2}(\nabla_iE_j+\nabla_jE_i),\quad H_{ij}={1\over 2}(\nabla_iH_j
+\nabla_jH_i)
\end{equation}
denote electric and magnetizing field gradients.  
While these quantities
are mathematically well-defined via Eq. \ref{eq:mno}, I am unable to
provide a good
physical picture.  The parameters $\gamma_{E1},\,
\gamma_{M1}$ are related to the classical Faraday rotation, wherein the
linear polarization of the photon passing longitudinally through a
magnetized medium exhibits a rotation due to the difference in index of
refraction for photons with circular polarization parallel and antiparallel
to the direction of magnetization.  However, I don't know how to go much farther
than this and will offer a bottle of fine German wine to anyone
who is able to provide me such a classical picture.

Again I will rely on Thomas Hemmert to present the theoretical situation, 
but on the experimental side, there exist as yet no direct polarized Compton
scattering measurements.  However, a global analysis of unpolarized Compton
data by the LEGS group has yielded the value\cite{ton}\footnote{Note here
that we have subtracted the pion pole contribution.}
\begin{equation}
\gamma_\pi=-\gamma_{E1}-\gamma_{M2}+\gamma_{E2}+\gamma_{M1}=
(15.7\pm 2.3\pm 2.8\pm 2.4)\times 10^{-4}\,{\rm fm}^4
\end{equation}
in disagreement with the theoretical prediction 
\begin{equation}
\gamma_\pi^{th}={\alpha g_A^2\over
96\pi^2F_\pi^2m_\pi^2}\left(4-({9\over 2}+\kappa_n){\pi m_\pi\over
M}\right)=3.3\times 10^{-4}\,{\rm fm}^4
\end{equation}
from ${\cal O}(p^4)$ heavy baryon chiral perturbation theory\cite{hbc}.  
However, a preliminary new value has been announced 
from the TAPS data
\begin{equation}
\gamma_\pi=(7.4\pm 2.3)\times 10^{-4}\,{\rm fm}^4
\end{equation}
which is in better agreement with theory.  The
other quantity about which much has been written is the forward
spin polarizability $\gamma_0$, which is given by the first moment
of the DGH sum rule
\begin{equation}
\gamma_0=-\gamma_{E1}-\gamma_{M2}-\gamma_{E2}-\gamma_{M1}=\int_{\omega_0}^\infty
{d\omega\over \omega^3}(\sigma_{3\over 2}(\omega)-\sigma_{1\over 2}(\omega))
\end{equation}
Drechsel has quoted perhaps the best current value of the
sum rule, based upon the MAID analysis,
\begin{equation}
\gamma_0=-0.75\times 10^{-4}\,{\rm fm}^4
\end{equation}
which is in reasonable agreement with previous determinations.

While at present we do not have direct experimental values for the four
spin-polarizabilities, it may be possible to extract them from future
$\vec{p}(\vec{\gamma},\gamma)p$ studies.  In the meantime, it has
been realized that they can be obtained using a dispersive analysis of
the Compton 
process\cite{pas}.  One assumes
that the Compton amplitudes $A_i$ can be
represented in terms of once subtracted dispersion relations at fixed t
\begin{equation}
A_i(\nu,t)=A_i^{\rm Born}(\nu,t)+(A_i(0,t)-A_i^{\rm Born}(0,t))
+{2\nu^2\over \pi}P\int_{\nu_{thr}}^\infty{{\rm Im}A_i(\nu',t)\over
\nu' ({\nu'}^2-\nu^2)}.
\end{equation}
Here Im$A_i(\nu',t)$ is evaluated using empirical photoproduction data
while the subtraction constant $A_i(0,t)-A_i^{\rm Born}(0,t)$ is
represented via use of t-channel dispersion relations
\begin{equation}
A_i(0,t)-A_i^{\rm Born}(0,t)=a_i+a_i^{t-pole}+
{t\over \pi}\left(\int_{4 m_\pi^2}^\infty
-\int_{-\infty}^{- 4 M m_\pi - 2 m_\pi^2} dt'
{{\rm Im}_tA_i(0,t')\over t'(t'-t)}\right)
\end{equation}
with Im$_tA_i$ evaluated using the contribution from the
$\pi\pi$ intermediate state.  In principle then there remain six
unknown subtraction constants $a_i$ to be determined
empirically.  However, in view of the limitations posed by the the data,
Drechsel et al. note that four of these quantities can be reasonably
assumed to obey unsubtracted forward dispersion relations, while the remaining
two---$\alpha_E-\beta_M$ and
$\gamma_\pi$---can be determined by an experimental fit.  Once this is
done the other spin polarizabilities may be
extracted using sum rules, as done above in the case of the forward
spin polarizability.  The results of this process
are given in Table 1, where they are compared to the numbers calculated in the 
${\cal O}(p^4)$ heavy baryon chiral perturbation theory calculation described by
Hemmert.

\begin{table}
\begin{center}
\begin{tabular}{ccc}
polarizability& HB$\chi$pt & Dispersive Evaluation\\
$\gamma_{E1}^p$&-1.8&-4.4\\
$\gamma_{M1}^p$&2.9&2.9\\
$\gamma_{E2}^p$&1.8&2.2\\
$\gamma_{M2}^p$&0.7&0.0
\end{tabular}
\caption{Calculated and "experimental" values for spin polarizabilities 
obtained via dispersion relations.  All are in units of $10^{-4}$ fm$^4$.}
\end{center}
\end{table}

It has been noted by Babusci et al.\cite{bab} and by Holstein et al.\cite{hol}
that one can
extend this analysis to include terms of ${\cal O}(\omega^4)$ in the Compton
amplitude by introducing higher order polarizabilities via
\begin{equation}
H_{eff}^{(4)}=-{1\over 2}4\pi\alpha_{E\nu}^p\dot{\vec{E}}^2
-{1\over 2}4\pi\beta_{M\nu}^p\dot{\vec{H}}^2
-{1\over 12}4\pi\alpha_{E2}^pE_{ij}^2
-{1\over 12}4\pi\beta_{M2}^p H_{ij}^2\label{eq:mmm}
\end{equation}
Likewise Holstein et al. have extended this to ${\cal O}(\omega^5)$ by
defining higher order spin-polarizabilities---
\begin{eqnarray}
H_{eff}^{(5)}&=&-{1\over 2}4\pi\left[\gamma_{E1\nu}^p\vec{\sigma}\cdot\dot{\vec{E}}
\times\ddot{\vec{E}}+\gamma_{M1\nu}^p\vec{\sigma}\cdot\dot{\vec{H}}\times
\ddot{\vec{H}}-2\gamma_{E2\nu}^p\sigma_i\dot{E}_{ij}\dot{H}_j+2\gamma_{M2\nu}^p
\sigma_i\dot{H}_{ij}\dot{E_j}\right.\nonumber\\
&+&\left.4\gamma_{ET}^p\epsilon_{ijk}\sigma_iE_{j\ell}\dot{E}_{k\ell}
+4\gamma_{MT}^p\epsilon_{ijk}\sigma_iH_{j\ell}\dot{H}_{k\ell}
-6\gamma_{E3}^p\sigma_iE_{ijk}H_{jk}+6\gamma_{M3}^p\sigma_iH_{ijk}E_{jk}\right]\nonumber\$
\quad
\end{eqnarray}
where
\begin{eqnarray}
{(E,H)}_{ijk}&=&{1\over 3}(\nabla_i\nabla_j{(E,H)}_k+\nabla_i\nabla_k{(E,H)}_j
+\nabla_j\nabla_k{(E,H)}_i)\nonumber\\
&-&{1\over
15}(\delta_{ij}\nabla^2(E,H)_k+\delta_{jk}
\nabla^2(E,H)_i+\delta_{ik}\nabla^2(E,H)_j)
\end{eqnarray}
are the (spherical) tensor gradients of the electric and magnetizing
fields.  Each of these new higher order polarizabilities
can be extracted via sum
rules from the Mainz dispersive analysis and results are given Table 2

Again Hemmert will discuss the details it is clear that 
the chiral perturbation theory description is remarkably
successful in describing all of these properties---the pion cloud
plays the dominant role in determining the polarizabilities.

\begin{table}
\begin{center}
\begin{tabular}{ccc}
polarizability&HB$\chi$pt&Dispersive value\\
$\alpha_{E\nu}^p$&2.4&-3.8\\
$\beta_{M\nu}^p$&7.5&9.3\\
$\alpha_{E2}^p$&22.1&29.3\\
$\beta_{M2}^p$&-9.5&-24.3\\
$\gamma_{E1\nu}^p$&-2.4&-3.4\\
$\gamma_{M1\nu}^p$&1.8&2.2\\
$\gamma_{E2\nu}^p$&1.6&1.3\\
$\gamma_{M2\nu}^p$&-0.1&-0.6
\end{tabular}
\caption{Calculated and "experimental" values for higher order 
polarizabilities obtained via dispersion relations.  Spin independent and
spin dependent polarizabilities are in units of $10^{-4}$ fm$^5$ and
$10^{-4}$ fm$^6$ respectively}
\end{center}
\end{table}

\section{VCS: Generalized Polarizabilities}
There has recently developed an interest in the process of virtual 
Compton scattering process by
which one can measure "generalized" (q-dependent) polarizabilities\cite{dho}.
In order
to understand the meaning of such quantities, recall that in ordinary electron
scattering measurement of the q-dependent charge form factor allows access,
via Fourier transform, to the nucleon charge {\it density}.  In an analogous
fashion measurement of a generalized polarizability such as $\alpha_E(q)$
permits one to determine the polarization density of the nucleon.  On the
experimental side this is an extremently challenging process because the
generalized polarizabilities can be determined only after (large) Bethe-Heitler
and Born diagram contributions have been subtracted.\footnote{Radiative
corrections are also substantial here.}  One then seeks a systematic
deviation---growing with $\omega'$---of the measured cross section from that
predicted with only Bethe-Heitler plus Born input in order to extract the
desired signal.  This has been achieved in a recent MAMI
experiment\cite{roche} and what
results is information on the two combinations
\begin{eqnarray}
P_{LL}-{1\over \epsilon}P_{TT}&=&a_0\alpha_E(q)-c_1\gamma_{M2}(q)
+c_2M^{M1-M1}(q)\nonumber\\
P_{LT}&=&b_0\beta_M(q)+c_3M^{C0-M1}(q)-c_4\gamma_{E2}(q)
\end{eqnarray}
where here the multipoles $M^{M1-M1},\,M^{C0-M1}$ have no RCS
analogs.  The extracted numbers are given in Table 3 together with values
calculated in various models as well as in HB$\chi$pt.  It is remarkable
that once again agreement with the simple chiral calculation is outstanding,
despite the fact that the measurement took place at $q$=0.6 GeV, where one
questions the validity of the chiral approach.

\begin{table}
\begin{center}
\begin{tabular}{c|c|c}
 &$P_{LL}-P_{TT}/\epsilon$&$P_{LT}$\\
\hline\\
expt.& $23.7\pm 2.2\pm 0.6\pm 4.3$& $-5.0\pm 0.8\pm 1.1\pm 1.4$\\
HB$\chi$pt\cite{hhks}&26.0&-5.3\\
L$\sigma$M\cite{lsm}&11.5&0.0\\
ELM\cite{elm}&5.9&-1.9\\
NRQM\cite{nrq}&11.1&-3.5
\end{tabular}
\caption{Measured and calculated values for generalized polarizabilities.
All are in units of GeV$^{-2}$.}
\end{center}
\end{table}

Fortunately, there is an additional experiment underway which should
help in this regard\cite{bates}.  A Bates measurement using the OOPS spectrometer
system has just begun taking data and will yield values of the
generalized polarizabilities at $q$=230 MeV, where the chiral
predictions should work.  Another interesting feature of this
measurement is that is it employs perpendicular kinematics (the lepton
and hadron planes are orthogonal) for which the Bethe-Heitler plus
Born ``background'' should be much less significant, allowing more
straightforward access to the desired generalized polarizabilities.

On the theoretical side, an extension of the
Mainz dispersive RCS analysis to the case of VCS has been done\cite{vdh}.
This calculation is not
as straightforward as it might appear, since replacement of a real photon
by a virtual one requires now {\it twelve} invariant amplitudes which are
functions of three variables, which may be taken as $\omega,\theta,Q^2$.
Analysis of the asymptotic dependence is correspondingly much more
complex.  In addition the subtraction ``constants'' are now functions
of $Q^2$ whose form must be assumed.
The calculation is still in progress but preliminary results
for four of the generalized polarizabilities compard to the chiral
predictions are available.   Agreement is satisfactory though not 
outstanding, but Hemmert has
argued that ${\cal O}(p^4)$ corrections will improve matters.

\section{VVCS: the GDH Sum Rule}

My final topic will be one that is not usually described in terms of
polarizabilities, but could be---that of doubly virtual Compton 
scattering.  This process is also virtual, in that it is not really
experimentally feasible.  Nevertheless, it is of great theoretical
interest, as we shall see.  In order to make things a simple as
possible we consider forward scattering with virtual photons of
identical $q^2$, for which the anti-symmetric component of the
scattering tensor can be written in the form\cite{jo}
\begin{eqnarray}
T_{\mu\nu}^{anti}&=&i\int d^4x e^{iq\cdot
x}<N_{p,s}|T(J_\mu(x)J_\nu(0)|N_{p,s}>_{anti}\nonumber\\
&=&-i\epsilon_{\mu\nu\alpha\beta}q^\alpha\left[S^\beta
S_1(\nu,Q^2)
+(\nu S_\beta-S-\cdot q{P^\beta\over M})S_2(\nu,Q^2)\right]
\end{eqnarray}
where, as usual, $\nu=p\cdot q/M$ and $S^\beta$ is the Pauli-Lubinski
spin vector.  By Regge or QCD arguments the sum rule for the non-Born
component of the form factor $S_1$ should converge, so we have
\begin{equation}
{1\over 4}S_1(0,Q^2)=\int_{\nu_0}^\infty{d\nu\over \nu}G_1(\nu,Q^2)
\equiv{1\over M^2}I(Q^2)={2M^2\over Q^2}\Gamma(Q^2)\label{eq:gdh}
\end{equation}  
What is interesting here is that in the real photon limit we have
\begin{equation}
I^{p,n}(Q^2)=-{1\over 4}\kappa^2_{p,n}
\end{equation}
where $\kappa_{p,n}$ is the anomalous moment so that we recognize 
Eq. \ref{eq:gdh} as being the Gerasimov-Drell-Hearn sum rule\cite{gdh}.  On the
other hand, in the limit as $Q^2\rightarrow \infty$ er jsbr
\begin{equation}
\Gamma(Q^2)^{p,n}\rightarrow \int_0^1 dxg_1(x,Q^2)\rightarrow
\pm{1\over 12}g_A+{1\over 36}g_8+{1\over 9}g_0
\end{equation}
where $g_A,g_8,g_0$ are the usual octet axial couplings, so that we 
recognize
\begin{equation}
\Gamma^p(Q^2)-\Gamma^n(Q^2)\stackrel{Q^2\rightarrow\infty}{\longrightarrow}
{1\over 6}g_A
\end{equation}
as being the Bjorken sum rule\cite{bj}.  Now, as argued by Ji and Osborne\cite{jo},  
QCD (operator product expansion)
methods can be used to extend the Bjorken sum rule down to values of
$Q^2$ as low as $\sim 0.5$ GeV$^2$ while chiral perturbation theory
methods can be used in order to go from the real photon point at
$Q^2=0$ up to higher values $\sim\leq 0.1$ GeV$^2$ so the challenge is
to connect the two pictures and work on this is ongoing.  In this
context it is interesting to make two comments:
\begin{itemize}
\item [i)] Much has been made of the possible discrepancy between the
theoretical and experimental values of the GDH sum rule for the 
difference of neutron and proton, while no one questions the
convergence of the Bjorken sum rule.  It seems to me that if one is
true then so must be the other.
\item [ii)] One can make a connection here with the spin-polarizability by
calculating the weighted integral
\begin{equation}
\int_{\nu_0}^\infty {d\nu\over \nu^3}G_1(\nu,Q^2)
\end{equation}   
but as yet this has not been looked at.
\end{itemize} 
There is much more to be said on both issues.

\section{Conclusion}
We have seen above that the subject of nucleon polarizabilities is now
a hot topic of research.  In the case of real Compton scattering, the
ordinary electric and magnetic polarizabilities are now well known and
the challenge is to extend such measurements to the higher order spin
polarizabilities.  In the mean time, dispersion relations combined
with sum rules can provide values for these numbers.  In the case of
virtual Compton scattering, the soon to be finalized generalized 
polarizbilities from the Mainz experiment are already challenging
theoretical models and we anxiously await results from ongoing
JLab and Bates experiments.  Finally, in the case of doubly virtual
Compton scattering, connections can be made between sacred sum rules
for both QCD and electromagnetic physics.  This is indeed an exciting
time for Compton scatterers.

\begin{center}
{\bf Acknowledgement}
\end{center}
It is a pleasure to acknowledge the hospitality of the Chiral Dynamics2000
organizers.  This work was supported in part by the National Science
Foundation.

\end{document}